\documentclass{ckm}                 % twocolumn proceedings style
\usepackage{txfonts}            % if you have it, to get Times family
\usepackage{bm}% bold math

\newcommand{\btou}{b\to u\ell\nu}
\newcommand{\btoc}{b\to c\ell\nu}
\newcommand{\Vub}{|V_{ub}|}
\newcommand{\pilnu}[1]{\pi^{#1}\ell\nu}
\newcommand{\rholnu}[1]{\rho^{#1}\ell\nu}
\newcommand{\omelnu}{\omega\ell\nu}
\newcommand{\etalnu}{\eta\ell\nu}

\newcounter{hwcount}
\newcommand{\homework}[1]{\stepcounter{hwcount}{\em Homework \#\thehwcount} \begin{quote} #1 \end{quote}}

\newcommand{\stackts}{\stackrel{\mathrm{tracks,}}{{\tiny\rm showers}}}
\newcommand{\mmiss}{M^2_{\mathrm{miss}}}
\newcommand{\Emiss}{E_{\mathrm{miss}}}
\newcommand{\pmiss}{\vec{p}_{\mathrm{miss}}}

\confname{Workshop on the CKM Unitarity Triangle, IPPP Durham, April
  2003}

\title{Towards Robust Averaging of Exclusive $B\to X_u\ell\nu$ Measurements}

\author{L Gibbons}
\address{Cornell University, Ithaca, NY, USA}

\begin{document}

\begin{abstract} I present a brief overview of the measurements of exclusive $B\to X_u\ell\nu$
transitions, with a focus on issues facing robust averaging of branching fractions and
$|V_{ub}|$ from current and anticipated measurements.
\end{abstract}

%% \maketitle needs to be after the author and address info and the
%% abstract... 
\maketitle

%% standard LaTeX from here on...

\section{Introduction}
With the foreseen improvements in theoretical techniques for form factor calculations,
measurement of exclusive $B\to X_u\ell\nu$ processes shows promise as the most robust route for
determination of $\Vub$.
The experiments BaBar, Belle and CLEO now have a variety of measurements, preliminary
or published, of  such processes. While the experiments employ a
 variety of strategies, the core techniques are similar and lead to potential correlations.
 With the current information available from the experiments, realistic evaluation of the correlations
 and, therefore, robust averaging of the results is very difficult.
This review surveys areas in which correlated systematic uncertainties
 are likely to exist for the measurements.  I pose a set of ``homework'' questions.  These
 can hopefully serve as a starting point for discussion among the experiments aimed at
 standardized evaluation of uncertainties where correlations are likely to exist.
 
 As discussed in any reference concerning exclusive $\btou$ transitions, the major theoretical uncertainty in the extraction of  $\Vub$ from these transitions arises from uncertainties in the hadronic form factors involved in the transitions (see, for example, \cite{cleo03}).  For electron and muon transitions, one form factor dominates transitions to final states with a single pseudoscalar meson, and three form factors dominate those to a single vector meson.  All of these form factors vary as a function of the momentum transfer $q^2 = M^2_{\ell\nu}$.  I will review mechanisms leading from uncertainties in the form factors to uncertainty in the branching fractions and in $\Vub$.   An excellent overview of the status of the theoretical work on the form factors can be found in the  proceedings of the previous CKM workshop \cite{Battaglia:2003in}.

Reference~\cite{Battaglia:2003in} also makes a valiant start towards an average of the various experimental results.  Better standardization among the experiments will be necessary to effect a more robust averaging procedure.

\section{Overview of measurements}

\subsection{General approach}

Table~\ref{tab:measurements} summarizes the measurements of exclusive $|V_{ub}|$ channels
that have been published or presented in preliminary form at conferences.  All
the measurements have made use of detector hermeticity to obtain an initial estimate the four 
momentum of the neutrino.  At BaBar, Belle and CLEO, the initial four momentum of the
$\Upsilon(4S)$ is known very well.  Since the detectors cover most of the total $4\pi$ solid angle,
the missing four momentum, defined via
\begin{eqnarray*}
\pmiss & = & \vec{p}_{\Upsilon(4S)} -  \sum_{\stackts}  \vec{p}_i\\ \nonumber
\Emiss & = & E_{\mathrm{\Upsilon(4S)}} -  \sum_{\stackts}  E_i \nonumber,
\end{eqnarray*}
provides a good estimate of the total four momentum of all undetected particles in the event.
For events with just one undetected neutrino from $B\to X_u\ell\nu$, the missing four momentum provides a good estimate of the neutrino's four momentum.

One of the challenges facing ``reconstruction'' of the neutrino in this fashion lies in appropriate selection of the tracks and showers used in the sums.  The BaBar '01 analysis, for example,
selects the subset of tracks $p_t > 100$ MeV$/c$, at least 12 hits in their drift chamber, and
consistent with coming from the origin.  The CLEO analyses  have optimized track selection for
hermeticity in two ways. For contributions to the missing momentum, CLEO attempts to identify one and only one reconstructed track per true particle from the origin (taking ``poorly measured'' in preference to ``not measured'').  Selection of tracks to be matched to hadronic showers in the calorimeter is performed separately.  For example, particles with transverse momentum low enough to "curl" within the detector may result in several reconstructed tracks.   CLEO groups these tracks and chooses the likeliest best representation of the original particle.  For particles that decay in flight or suffer a hard scatter or a hadronic interaction, CLEO identifies the inner and outer tracks for the two different purposes.  Similarly, the experiments have a range of selection criteria of showers in the electromagnetic calorimeter to separate true photons from showers originating with interactions of charged hadrons.
\begin{table}
  \centering 
  \caption{Published and preliminary $B\to X_u\ell\nu$ measurements.  The years correspond to year of publication, year of submission for publication, or year of presentation for identification purposes and are {\em not} indicative of the intellectual history.  For example, the preliminary version of CLEO '03 and of Belle '02 appeared simultaneously.}
  \label{tab:measurements}
\begin{tabular}{ll}
\hline\hline
Measurement & Modes studied \\ \hline
Belle 2003 \cite{belle03}  &  $\omelnu$ \\
CLEO 2003 \cite{cleo03}   & $\pilnu{\pm}$, $\pilnu{0}$,$\rholnu{\pm}$, $\rholnu{0}$,$\omelnu$, $\etalnu$ \\
BELLE 2002\cite{belle02}   & $\pilnu{\pm}$, $\rholnu{0}$  \\
BaBar 2001 \cite{babar01}  & $\rholnu{\pm}$, $\rholnu{0}$, $\omelnu$  \\
CLEO 2000 \cite{cleo00}  & $\rholnu{\pm}$, $\rholnu{0}$, $\omelnu$  \\
CLEO 1996 \cite{cleo96} & $\pilnu{\pm}$, $\pilnu{0}$,$\rholnu{\pm}$, $\rholnu{0}$,$\omelnu$ \\
\hline
\end{tabular}
\end{table}

\begin{figure}
\begin{center}
\hbox to\hsize{\hss
\includegraphics[height=3.5 in]{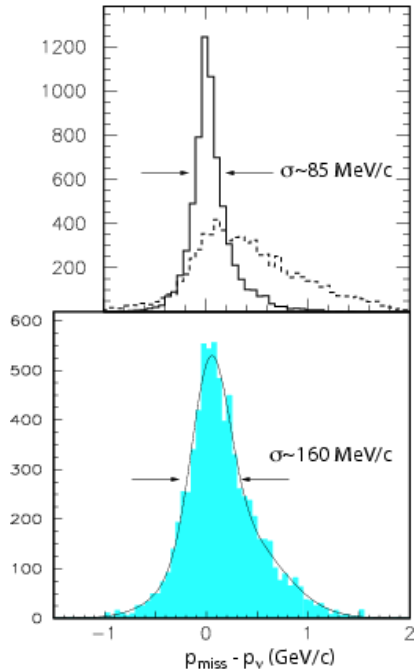}
\hss}
\caption{$|\pmiss|-|\vec{p}_\nu|$ for simulated $B\to X_u\ell\nu$.  Top: CLEO '03 distributions without (solid) and with (dashed) missing extra neutrals.  No strict event selection criteria (see text) have been applied.  Bottom: Total Belle '03 distribution after application of the strict even selection criteria.  In both cases, the core resolution has been indicated.}
\label{fig:pmiss}
\end{center}
\end{figure}

The missing momentum yields an accurate representation of a missing neutrino from a semileptonic decay only when that is the sole particle that has not been detected.  In fact,
the largest background contribution in these analyses, as in the inclusive $\btou$ analyses, arises from events containing a $\btoc$ decay with at least one neutral particle missing in addition to the neutrino (see, for example, reference \cite{cleo96}).   A number of the analyses introduce strict event selection criteria to bias against events containing multiple missing particles.  For example, the CLEO '96, '03 and the Belle '01, '03 analyses require precisely one identified charged lepton (additional implying extra missing neutrinos), and a small reconstructed net charge to the event (nonzero implying at least one missing or double-counted particle).  In addition, those analyses require the missing momentum to be consistent with a neutrino.  The Belle analyses require a
small $\mmiss$, while the CLEO analyses require a small $\mmiss/\Emiss$.  Since the $\Emiss$ resolution is considerably worse than the $\pmiss$ resolution (see, for example, reference~\cite{cleo03}), the latter requirement is roughly constant in $\Emiss$ resolution.  
The missing momentum resolution is shown for CLEO and Belle in Figure~\ref{fig:pmiss}.

The CLEO '00 and BABAR '01 $\rho\ell\nu$ analyses do not apply these strict event selection criteria.  As a result, the analyses have a much higher  signal efficiency.  That efficiency, however, comes with a much fiercer $\btoc$ background.  Hence both of those analyses are primarily sensitive to their signal in the region $p_\ell>2.3$ GeV$/c$, which lies beyond the kinematic endpoint for the $\btoc$ transitions.  As we will discuss below, this approach shifts systematic uncertainties from detector-related uncertainties to signal form factor shape uncertainties.

As the $B$ factory datasets continue to grow, the exclusive $\btou$ analyses will benefit from use of fully reconstructed (or ``annealed'') samples used in the recent inclusive analyses (see talk by F.~Muheim \cite{inclusive_review}).  Such an approach should result in a significant reduction in background, allowing for selection criteria that can be made more uniform over phase space.  As a result, systematic uncertainties both from detector and background modeling and from form factor uncertainties will be reduced.

\begin{figure}[b!]
\hbox to\hsize{\hss
\includegraphics[width=\hsize]{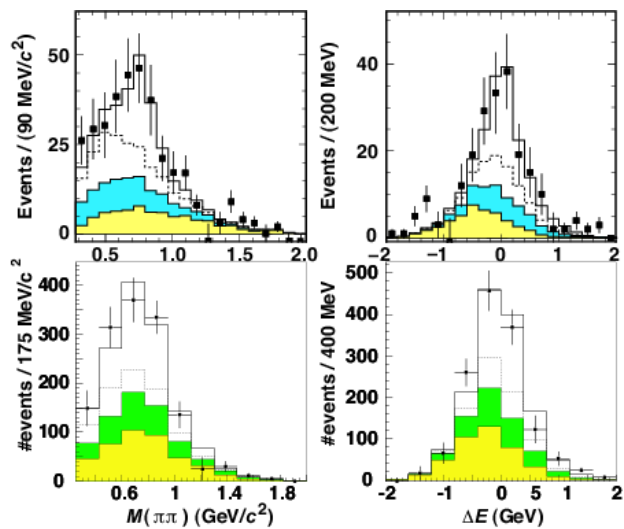}
\hss}
\caption{Projections of the fit for signal yields in the CLEO '00 (top) and BaBar '01 (bottom)
analysis onto $m_{\pi\pi}$ (left) and $\Delta E$ (right) for events with $p_\ell>2.3$ GeV/$c$. The
points are continuum-subtracted data.  The histograms are, from bottom to top, $\btoc$, 
other $\btou$, cross-feed between the $\rholnu{\ }$ modes and $\rholnu{\ }$ signal.}
\label{fig:lange_fits}
\end{figure}

After event selection, the analyses use a neutrino four momentum given by $p_\nu = (|\pmiss|,\pmiss)$ because the missing momentum resolution is considerably better than the missing energy resolution.  The experiments then consider a variety of kinematic variables related to a single $\btou$ decay.  The analyses either employ these for further background suppression or fit some combination of them for extraction of the signal yields.   The variables
\begin{eqnarray}
\Delta E & = & E_m + E_\ell + E_\nu - E_{\mathrm{beam}} \\
M_{\mathrm{beam}} & = & (E_\mathrm{beam}^2 - |\vec{p}_m+\vec{p}_\ell+\vec{p}_\nu)^{1/2} 
\end{eqnarray}
characterize energy conservation ($\Delta E=0$) and momentum conservation ($M_{\mathrm{beam}}=M_B$) in the decay, and therefore involve the ``measured'' $p_\nu$.
Alternatively, one can calculate the angle between the meson+charged lepton system ($Y$) and the B meson, without recourse to $p_\nu$, via
\begin{equation}
\cos \theta_{BY}=\frac{2 E_B E_{Y} - (M_B^2  + M_{Y}^2) c^4} 
                      {2 |\vec{p}_B| |\vec{p}_{Y}| c^2}.
\end{equation}
The variables $\cos\theta_{BY}$ and $M_{\mathrm{beam}} $ are strongly correlated:  a candidate combination cannot have $M_{\mathrm{beam}}$ consistent with the $B$ meson mass without $\cos\theta_{BY}$ being within (or close to) its physical domain.  All but the CLEO '96 and CLEO '03 analyses require an approximately physical $\cos\theta_{BY}$.

\subsection{Extraction of $\btou$ yields}

\begin{figure}[b!]
\hbox to\hsize{\hss
\includegraphics[width=\hsize]{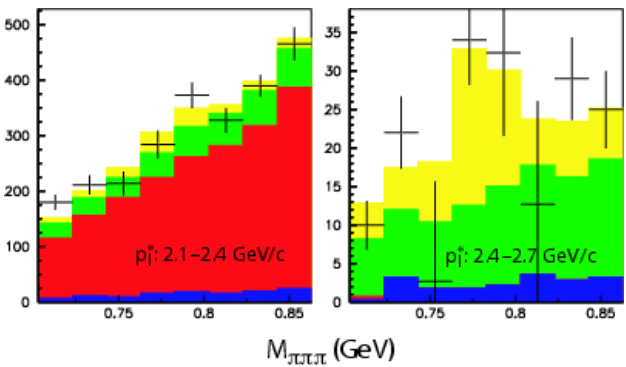}
\hss}
\caption{The $m_{\pi\pi\pi}$ distributions in the two $p_\ell$ intervals indicated from the Belle '03 analysis.  The momentum intervals are examined in the $\Upsilon(4S)$ rest frame.  The points are on-resonance data. The fit components are, from the bottom up, continuum+fake (blue), $\btoc$ (red), other $\btou$ (green), and signal $\omelnu$ (yellow).}
\label{fig:belle_omega}
\end{figure}

To extract yields and thus obtain branching fractions, all experiments perform a multicomponent fit with Monte Carlo or data estimates of the $\btoc$ backgrounds, $\btou$ signal and backgrounds (cross-feed among modes considered or feed down from modes not considered), and continuum backgrounds.  The fits employ isospin (and quark model) constraints
\begin{eqnarray}
\Gamma(B^0\to \pi^-\ell^+\nu) & = & 2\Gamma(B^+\to \pi^0\ell^+\nu) \\ \nonumber
\Gamma(B^0\to \rho^-\ell^+\nu) & = & 2\Gamma(B^+\to \rho^0\ell^+\nu) \\ \nonumber
                                                         & \approx & 2\Gamma(B^+\to \omega\ell^+\nu) \label{eq:isospin}
\end{eqnarray}
when both the charged and neutral $B$ decay modes are measured in an analysis. 

The CLEO '00 and BaBar '01 $\rho\ell\nu$ analyses fit the $m_{\pi\pi}$ versus $\Delta E$ distribution in coarse bins of $p_\ell$.  Again, while these are relatively high efficiency analyses, they are primarily sensitive to signal in the region $p_\ell>2.3$ GeV/$c$
(in the $\Upsilon(4S)$ rest frame). The CLEO '00 analysis made the first determination of $d\Gamma/dq^2$ for any exclusive $\btou$ transition, albeit with large uncertainties.   Figure~\ref{fig:lange_fits} shows the fit results projected onto  $m_{\pi\pi}$ and $\Delta E$ distributions for events with $p_\ell>2.3$ GeV/$c$. 

The Belle analyses use a variety of approaches for extracting the yield.  The Belle '02 $\pilnu{+}$ analysis fits $\Delta E$ versus $p_\ell$.   After obtaining background and signal component normalizations from the fit, Belle has made a preliminary attempt to obtain $d\Gamma/dq^2$ by subtracting the background components from the reconstructed $d\Gamma/dq^2$ distribution and then correcting for resolution effects via an efficiency matrix. The Belle '02 $\rholnu{0}$ analysis requires $2.0\le p_\ell < 2.8$ GeV$/c$, then extracts yields using a fit to the $\Delta E$ versus
$m_{\pi\pi}$ distribution.  

Belle has observed  the $\omelnu$ transition  with their Belle '03 analysis.  After requiring $\Delta E$ and $M_{\mathrm{beam}}$ to be consistent with their signal, Belle performs a fit to $m_{\pi\pi\pi}$ versus $p_\ell$.  As in the CLEO '00 and BaBar '01 analyses, the data are coarsely binned in $p_\ell$ (in the $\Upsilon(4S)$ rest frame).  Similarly to those analyses, the analysis is primarily sensitive to signal with $p_\ell>2.4$ GeV$/c$.  Figure~\ref{fig:belle_omega} shows the $m_{\pi\pi\pi}$ distributions in two of the $p_\ell$ ranges.

\begin{figure}
\hbox to\hsize{\hss
\includegraphics[width=\hsize]{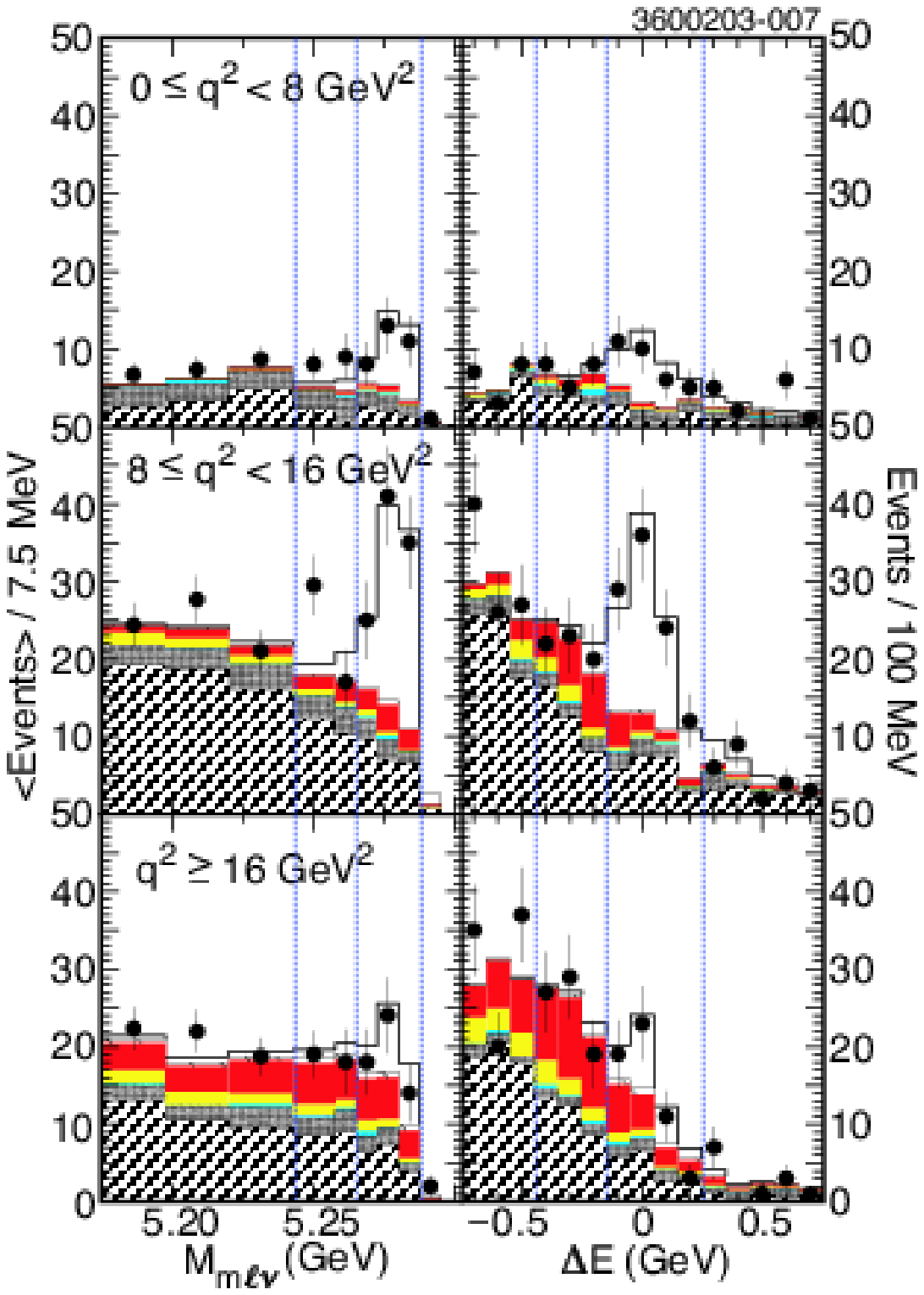}
\hss}
\caption{The $M_{\mathrm{beam}}$ and $\Delta E$ distributions in the other's signal region in each of the three $q^2$ intervals from the CLEO '03 fit.  The points are on-resonance data.  The histograms, from bottom to top, are $\btoc$ 45$^\circ$ hatch), continuum (grey), fake leptons (light blue), non-signal ($\pilnu{\ }$, $\rholnu{\ }$, $\omelnu$ or $\etalnu$) $\btou$ (yellow), cross-feed from other signal modes (red), cross-feed between $\pilnu{\ }$ modes (135$^\circ$ hatch), and $\pilnu{\ }$ signal (open).}
\label{fig:cleo_pi}
\end{figure}

\begin{figure*}
\hbox to\hsize{\hss 
\includegraphics[width=0.45\hsize]{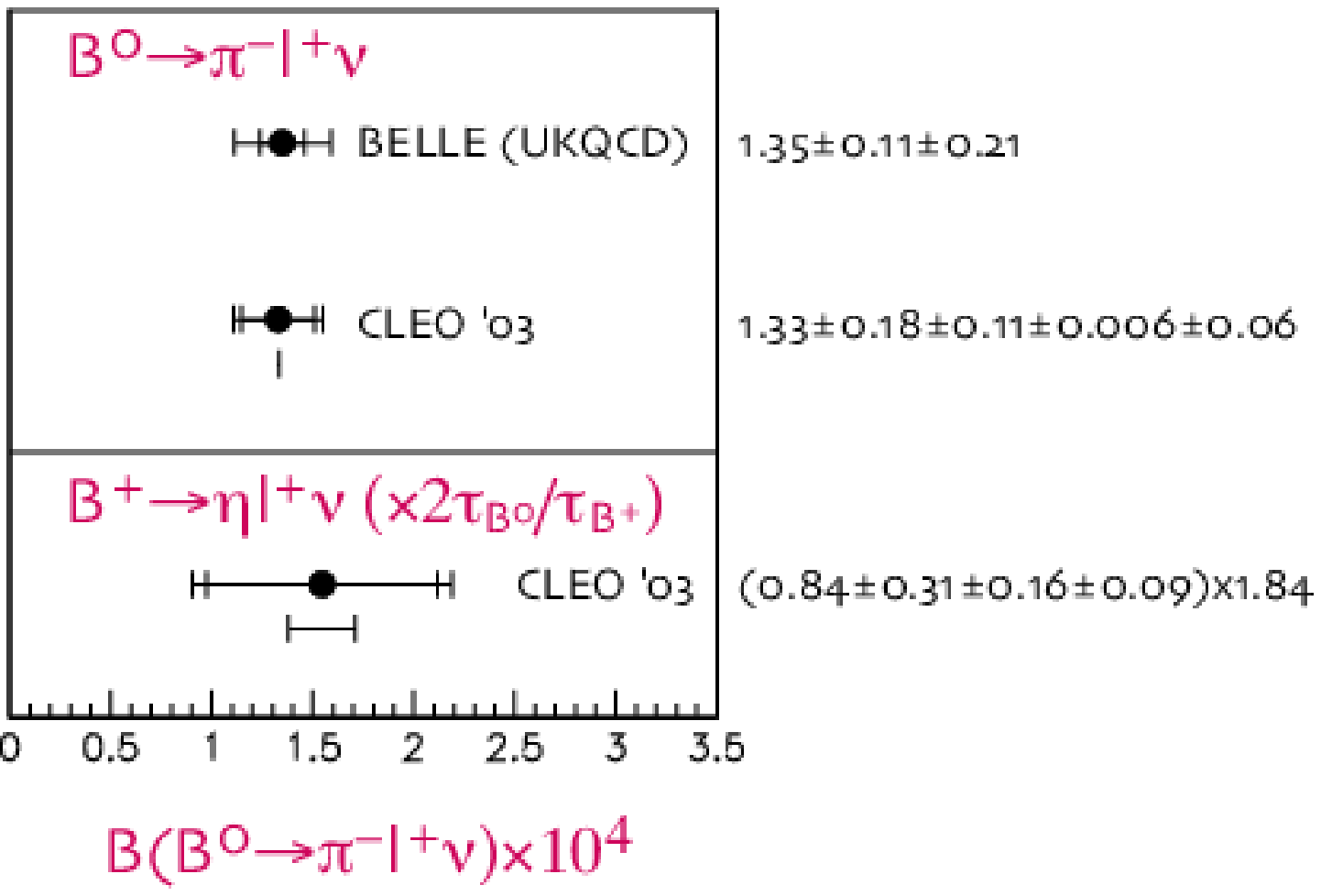}
\hfill
\includegraphics[width=0.45\hsize]{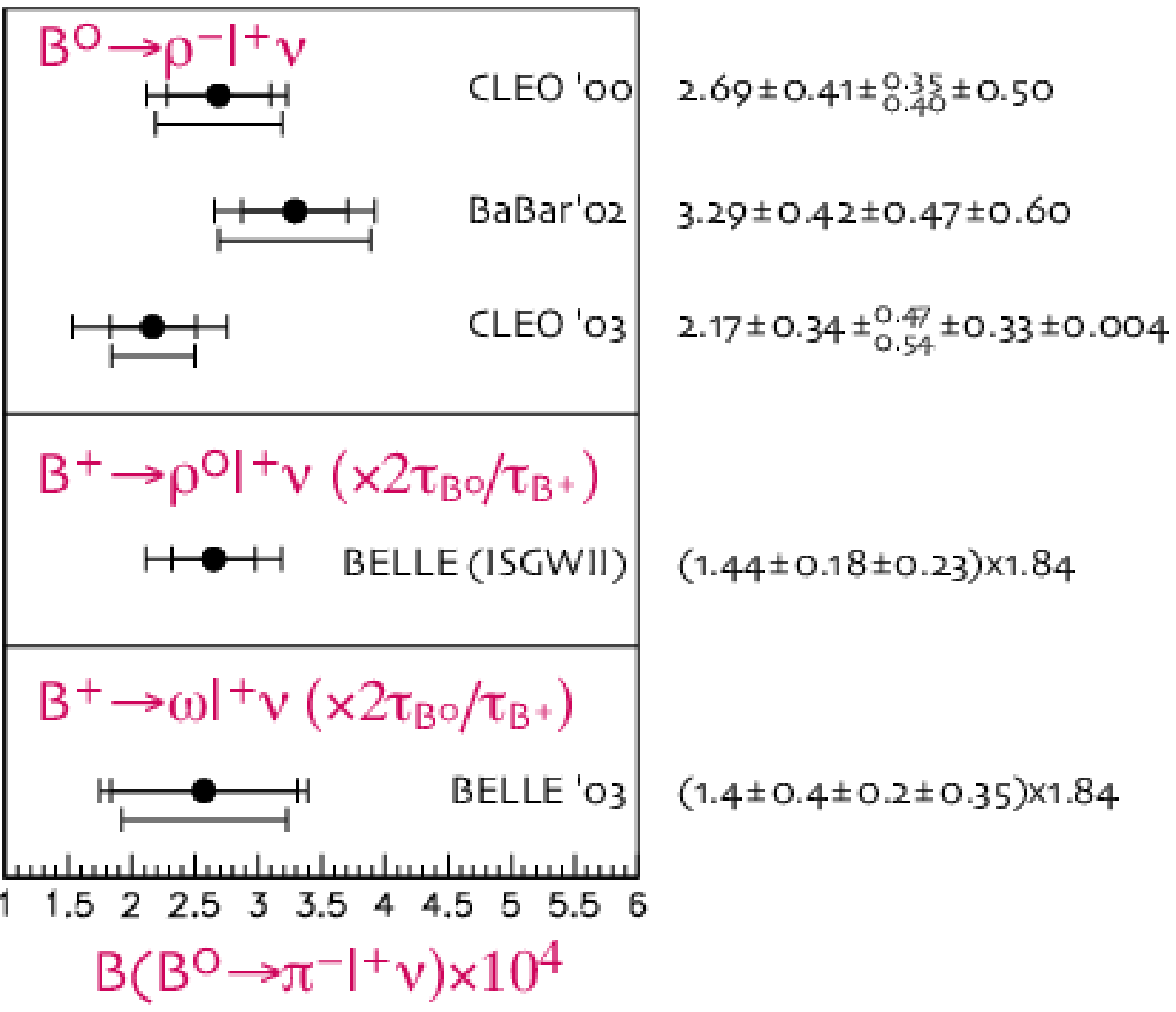} 
\hss}
\caption{Summary of the branching fraction measurements of exclusive $B\to X_u\ell\nu$ modes.
Left: modes to a pseudoscalar meson. Right: modes to a vector meson.  Note that all the $B^0$ branching fractions listed, except BELLE's, are obtained by including data from both the $B^0$ and corresponding $B^+$ decay modes with partial widths constrained  via isospin and quark model predictions (Eq.~\protect\ref{eq:isospin}).  For $B^+$ decay modes measured independently, the branching fractions are plotted with a $2\tau_{B^0}/\tau_{B^+}=1.84$ scale factor to allow for direct comparison with the $B^0$ branching fractions.  The experimental statistical and systematic uncertainties are indicated on each measurement.  The theoretical uncertainty from the form factor $q^2$--dependence, when evaluated, is indicated underneath the measurement.  Otherwise, the form factor calculation used is indicated in parentheses. To allow for direct comparison among experiments, the value quoted for the theoretical uncertainty is $1/2$ the spread obtained from the set of form factors considered. This should not be interpreted as an endorsement of this method for determination of that uncertainty.}
\label{fig:branching_fractions}
\end{figure*}

The CLEO '03 analysis design minimizes model dependence as much as possible.  The $p_\ell$ requirements (in the $\Upsilon(4S)$ rest frame) are relatively loose: $p_\ell>1.0$ GeV$/c$ for pseudoscalar modes and $p_\ell>1.5$ GeV$/c$ for vector modes.  The yields are extracted from a fit to $\Delta E$ versus $M_{\mathrm{beam}}$ distributions.  The vector modes are also binned in the reconstructed $\pi\pi$ or $3\pi$ mass. Unlike in the above analyses (and in the CLEO '96 analysis, which it supersedes), the signal rates are allowed to float independently in separate $q^2$ intervals.  As discussed  in the following section, this approach significantly reduces the uncertainty arising from the $q^2$--dependence of the form factors.  The total branching fraction is obtained by summing the three individual rates obtained in the fit.  By contrast, all other  analyses obtain the branching fraction by integrating over the $q^2$ distribution predicted by theory+simulation and fitting the data for a single normalization.  Figure~\ref{fig:cleo_pi} shows the  $M_{\mathrm{beam}}$ and the $\Delta E$ distributions for the $\pilnu{\ }$ mode in the other's signal region.

The branching fractions from all of the above measurements are summarized in Figure~\ref{fig:branching_fractions}.  The measurements agree quite well within their listed
uncertainties, though the extent of correlation among the measurements has not been evaluated.  Within the large experimental uncertainties, the independently measured $B^+$ rates  agree well with isospin and quark model predictions (Eq.~\ref{eq:isospin}).

\section{Averaging issues} 

\begin{table*}
  \centering 
 \caption{Systematic contributions (\%) from the Babar '01 analysis (left) and the CLEO '03
 analysis (right), except for contributions from signal mode form factor uncertainties.}
 \label{tab:allsyst}
\begin{tabular}{l|c} \hline \hline
Contribution                               & $\delta{\cal B}_{\rho}/{\cal B}_{\rho}$ (\%) \\ 
\hline
Tracking efficiency                        & $\pm 5$\\
Tracking resolution                        & $\pm 1$\\
$\pi^0$ efficiency                         & $\pm 5$\\
$\pi^0$ energy scale                       & $\pm 3$\\
\hline
$b\rightarrow c e \nu$ bkg.    & $+1.4,-1.7$ \\
Resonant $b\rightarrow u e \nu$ bkg. & $+6,-4$ \\
Non-resonant $b\rightarrow u e \nu$ bkg.  & $\pm 9$\\
\hline
$B$ lifetime                               & $\pm 1$ \\
Number of $B\bar{B}$ pairs                 & $\pm 1.6$ \\
Misidentified electrons                    & $<\pm 1$ \\
Electron efficiency                        & $\pm 2$ \\
${\cal B}(\Upsilon(4S)\rightarrow B^+B^-)/{\cal B}(\Upsilon(4S)\rightarrow B^0\bar{B}^0)$ & $<\pm 1$ \\
Isospin and quark model symmetries         & $<\pm 1$ \\
Fit method                                 & $+4,-6$ \\ \hline
Total systematic uncertainty               & $\pm 14.4$ \\
\hline \hline
\end{tabular}
\hfill
\begin{tabular}{rccc}  \hline\hline
Systematic & $\pi\ell\nu$ & $\rho(\omega)\ell\nu$ & $\eta\ell\nu$ \\  \hline
hermeticity                                &6.8 &  18.7&   17.3 \\ 
$B\to D/D^{*}/D^{**}/D^{\rm NR}X\ell\nu$    &1.7 &   2.0&    5.5\\
$B\to X_u\ell\nu$`w feed down                 &0.5 &   8.3&    1.6\\
Continuum smoothing                         &1.0 &   3.0&    2.0\\
Fakes                                       &3.0 &    3.0&   3.0\\
Lepton ID                                   &2.0 &    2.0&   2.0\\
$f_{+-}/f_{00}$                             &2.4 &    0.0&   4.1\\
$\tau_{B^+}/\tau_{B^0}$                     &0.2 &    2.1&   1.4\\
Isospin                                     &0.0 &    2.4&   0.1\\
Luminosity                                  &2.0 &    2.0&   2.0 \\ \hline
Upper Total                              &{\bf 8.6} &{\bf 21.4}& {\bf 19.3}\\ \hline
Non Resonant                                &-- &   -13&        \\ \hline
Lower Total                               &{\bf 8.6} &{\bf 25.1}& {\bf 19.3} \\ \hline\hline
 & & & \\
 & & & \\
\end{tabular}
\end{table*}

To perform a robust average of branching fractions and of $\Vub$ extracted from the measured rates, careful consideration of correlated uncertainties will be necessary.  As many of the results are quite new, the experiments have not  yet done the work examining the correlations.  As a result, I do not present an average in these proceedings.  However, I summarize below the primary issues that the experiments must consider to effect a realistic average.  Some of the effects may appear small on the scale of the statistical uncertainty of the current individual measurements. I anticipate, however, that a variety of results with improved precision will become available, and that averages with a statistical precision of 5\% or better may finally become possible.  We should anticipate this scenario, and begin examination of systematic contributions that are significant on this final scale, particularly if they may be largely correlated among the measurements.  

Table~\ref{tab:allsyst} provides the systematics breakdown from the two most recent published analyses.  Clearly
hermeticity--related issues, modeling of the generic $B\to X_u\ell\nu$ background, and limiting the nonresonant
$\pi\pi\ell\nu$ background contribute substantially in the measurements, particularly in the $\rho\ell\nu$ mode.
What is not apparent in the table is that treatment of the uncertainties in the inclusive rate and modeling can
exacerbate the hermeticity--related uncertainties.  Changes of the detector response, for example, affect the
best-fit normalizations for the  $B\to X_u\ell\nu$ background, which in turn can cause a significantly larger 
variation in signal shape than obtained with a fixed  $B\to X_u\ell\nu$ normalization.  The following subsections
discuss the systematic issues and potentially large correlations for the important categories.

\subsection{Systematics related to technique}

\begin{table}
  \centering 
 \caption{Systematics associated with the use of hermeticity in the CLEO '03 analysis.  The individual bold--faced items correspond to systematic effects that are likely to be at least partly correlated among the various experiments.}
 \label{tab:hermsyst}
 \vspace{5 pt}
 \begin{tabular}{cccccccccc}
\hline\hline
variation  & $\pi^-\ell^+\nu$ & $\rho^-\ell^+\nu$ & $\eta\ell\nu$ \\ \hline
$\gamma$ eff.      &  2.6 &  11.1  & 5.7  \\
$\gamma$ resol.  &  4.1 &  2.9  & 9.6  \\
$\mathbf{K_L}$ \textbf{showering}      &  1.3 &  6.0  &  2.7  \\
Particle ID              &  1.9 &  8.2   & 0.2  \\
\textbf{Splitoff rejection}    &  1.5 &  1.2   & 5.5  \\
\textbf{track eff.}                  &  3.7 &  8.6  & 9.5  \\
track resol.             &  1.0 &  6.2  & 0.9  \\
\textbf{splitoff sim.}             &  0.4 &  1.0  & 6.0  \\
$\mathbf{K_L}$ \textbf{production} &  0.2 &  0.1  & 0.1  \\
$\mathbf{\nu}$ \textbf{production}   &  0.5 &  0.6  & 2.9  \\ \hline
{\bf Total}                 & 6.8  & 18.7 & 17.3  \\
\hline\hline
\end{tabular}
\end{table}

All of the analyses performed to date rely on detector hermeticity to provide an estimate of the neutrino momentum and, as a result, more robust background suppression.  The price to be paid is an enhancement of the systematic uncertainties from physics and detector simulation.   A systematic uncertainty in the tracking efficiency, for example, acts coherently over all charged particles in an event.  Since there are typically about ten charged particles per $B\bar{B}$ event at the $\Upsilon(4S)$), single track effects can get magnified dramatically.   {\em A priori} it is difficult to predict how correlated the systematics will be and how they will vary from experiment to experiment.  The systematic effects tend to affect both background levels and efficiency, so the effects could vary quite significantly among the measurements depending on the optimization of efficiency versus background level. 

The hermeticity--related systematic categories considered by the CLEO '03 analysis are tabulated in Table~\ref{tab:hermsyst}.  Because detector simulation in the experiments tend to rely on a common GEANT base, or even more fundamentally, on a common pool of cross section measurements, many of the categories are likely to be at least partially correlated among the experiments.  CLEO has found, for example, measurable differences between data and simulation for hadronic interactions of kaons in its Cesium--Iodide (CsI) calorimeter.  These studies provide the input for the $K_L$ shower simulation listed in the table.  While BaBar and Belle may be less susceptible to these effects because of their instrumented flux returns, this could remain a sizable, strongly correlated systematic.  CLEO has also found that data and simulation disagree on the energy deposited in charged hadronic showers at distances too large to veto using proximity to a projected track.  The resulting systematic contributions in the production of such ``splitoffs'' and in the variables used to reject them appear, fortunately, to be rather small.  They are, however, also likely to be significantly correlated among experiments, so could result in a non-negligible contribution to an experimental average.

The CLEO tracking system and its associated uncertainties have been intensively studied at CLEO. As a result, the individual track--finding efficiency has been limited in studies at the $0.5\%$ level.  The CLEO '03 analysis has used $0.75\%$ per track in Table~\ref{tab:hermsyst} to be conservative.  Even at this small rate, the track efficiency uncertainty becomes sizable when coherently amplified by all tracks in the event.   At $0.5\%$ level, the uncertainty in the tracking
efficiency studies that provided that limit are similar to the uncertainty in total number of interaction lengths in the tracking system because of uncertainties in the underlying cross sections.

At a smaller level in the CLEO analysis are the uncertainties in the $K_L$ production spectrum and rate and in the $\nu$ production spectrum and rate (particularly from the secondary production process $b\to c\to s\ell\nu$) from the $\Upsilon(4S)$.   The production models tend to be strongly correlated among experiments through common event generators, so these categories still merit some attention. 

Other systematics in the list may also be correlated, such as the photon resolution resulting from a common model of electromagnetic showers.  Typically in measurements many of these effects are small.  However, with the amplification from the use of hermeticity, the experiments should evaluate with some care the levels of uncertainty and the levels of correlation among experiments for a broad variety of detector effects.

\homework{What is the complete set of systematics related to generic production model and detector simulation that are significant at the few percent level for the exclusive $\btou$ analyses?}
\homework{What is the level of correlation among the experiments for these different systematic effects?}

\subsection{Anatomy of rate measurement dependence on form factors}

The literature now teems with calculations of the form factors involved in the $\pilnu{\ }$ and 
$\rholnu{\ }$ transitions.  For a relatively recent survey, see the references within reference~\cite{cleo03}.  This review will limit itself to those considered within the analyses to date.
These include a variety of  lattice QCD (LQCD) calculations from APE (APE '00) \cite{Abada:2000ty}, FNAL (FNAL '01) \cite{El-Khadra:2001rv}, JLQCD (JLQCD '01) \cite{Aoki:2001rd},  and UKQCD (UKQCD '98 and UKQCD '99) \cite{DelDebbio:1997kr,Bowler:1999xn}
 and light-cone sum rules (LCSR) calculations from Ball and collaborators (Ball '98 and Ball '01) \cite{Ball:1998kk,Ball:2001fp} and Khodjamirian and collaborators (KRWWY) \cite{Khodjamirian:2000ds}.  Both approaches currently have uncertainties in the $15\%$ to $20\%$ range.  Also incorporated into various analyses are the quark model calculations of Isgur {\em et al}  (ISGW~II) \cite{Scora:1995ty}, Feldman and Kroll (SPD) \cite{Feldmann:1999sm}, and the relativistic quark model calculation of Melikhov and Stech \cite{Melikhov:2000yu}.   As they are purely models, it is difficult to assign a meaningful uncertainty to the quark model calculations. Finally, the analysis of Ligeti and Wise \cite{Ligeti:1995yz} based on study of heavy quark symmetry and measured $D\to K^*\ell\nu$ form factors is often considered.
 
 \begin{figure}
\hbox to\hsize{\hss
\includegraphics[width=\hsize]{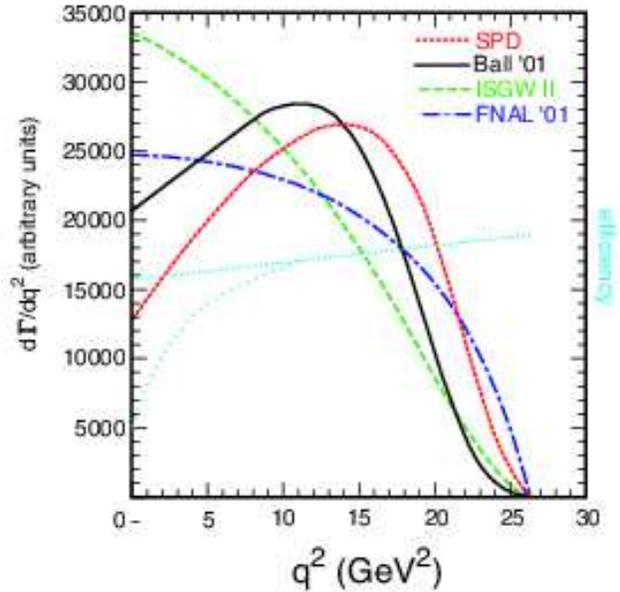}
\hss}
\caption{The $d\Gamma(B\to\pilnu{\ }/dq^2$ distributions arising from a variety of form factor calculations.  The calculations were chosen to indicate the significant variation in the predicted distribution that persists in the current literature.  Also shown (light blue) are rough representations of the efficiency variation that can result from a variety of background suppression requirements (see text).}
\label{fig:pi_models}
\end{figure}

A rate measurement suffers from significant theoretical uncertainties when the experimental efficiency varies significantly over the variables on which the form factor, or combination of form factors, depend.  To clarify the problem, consider the $\pilnu{\ }$ transition, for which a variety of form factor calculations are shown in Figure~\ref{fig:pi_models}.  The models were selected to indicate the span of $q^2$ dependence that arises in the literature: a significant variation in the case of $\pilnu{\ }$.  Since there is a single form factor that dominates the dynamics for this mode (in the limit of massless leptons), the $d\Gamma/dq^2$ distribution contains all of the dynamical information.  If  an analysis (a) integrates over a broad $q^2$ interval and (b) imposes selection criteria that cause a significant efficiency variation, then a significant variation in the form factors will cause a significant variation in the weighting of low--efficiency versus high--efficiency $q^2$ regions.  As a result, the overall efficiency prediction will vary dramatically, leading to significant theoretical uncertainty.

A variety of effects can lead to a $q^2$--dependent efficiency.  All analyses, for example, require a minimum lepton momentum, which results in the reconstruction efficiency falling as $q^2$ with $q^2$ (sloped light blue line in Figure~\ref{fig:pi_models}.  The higher the minimum momentum, the steeper the efficiency slope becomes.

\begin{figure}
\hbox to\hsize{\hss
\includegraphics[width=\hsize]{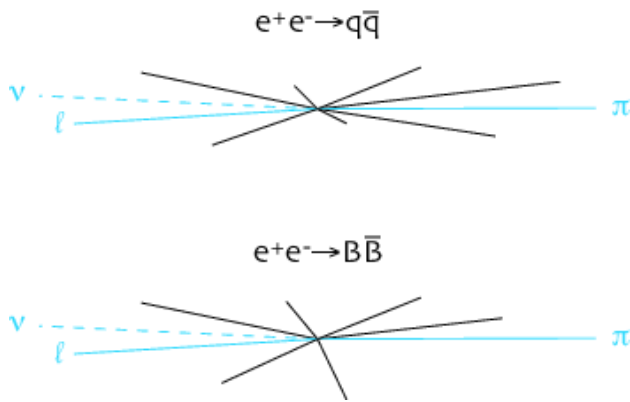}
\hss}
\caption{Top: sketch of a typical continuum process that leads to background contributions to a $B\to\pilnu{\ }$ at the $\Upsilon(4S)$.  Bottom: sketch of the topology of a $B\bar{B}$ event with a $B\to\pilnu{\ }$ decay at low $q^2$.}
\label{fig:topology}
\end{figure}

A more subtle situation can arise from methods of continuum background suppression.  The continuum background naturally reconstructs at low $q^2$.  To achieve a large beam--constrained mass (near $M_B$),  the kinematics favor the selection of a meson (in this case a pion) from one jet and a lepton pair from a semileptonic decay in the opposite jet (see Figure~\ref{fig:topology}). Interpreted as an $\Upsilon(4S)$ event, the $\ell\nu$ pair has a small invariant mass, and hence the $\pilnu{\ }$ candidate decay appears to have low $q^2$.   This topology, unfortunately, mimics that of a true $\Upsilon(4S)$ event containing a $B\to\pilnu{\ }$ decay at low $q^2$ (Figure~\ref{fig:topology}).  Continuum background suppression techniques based on variables that characterize the event {\em shape} generally introduce a significant bias into signal reconstruction efficiency versus $q^2$.  Requiring the ratio of the second to zeroth Fox--Wolfram moments, $R2$, to be small (spherical topology) introduces such a bias.  Variables characterizing the momentum flow relative to the lepton momentum direction, while providing powerful continuum suppression, severely bias the reconstruction efficiency.  The efficiency curve that falls steeply at low $q^2$ in Figure~\ref{fig:pi_models} characterizes the effect of lepton--based suppression.  Fortunately, continuum suppression can be realized without introducing any bias into the reconstruction efficiency.  The angle between the thrust or sphericity axes for the candidate event and the remainder of the event provides such suppression (see, for example, Ref.~\cite{lkgAnnRev}).

The same considerations hold for other $\btou$ decay modes.  Decays to a vector meson cannot, of course, be characterized solely by the $d\Gamma/dq^2$ distribution because of the interference among the three form factors involved in those decays.

Analyses can, of course, be optimized to reduce effects from uncertainties in the form factors. Obviously, the looser the selection criteria on lepton momentum and event shape variables that bias the efficiency versus $q^2$, the smaller the bias.  One can further reduce the form factor uncertainty via extraction of independent rates in separate $q^2$ intervals.  The form factors are then only required to provide relative rates over each restricted $q^2$ interval, and not between the different $q^2$ intervals.  As  a bonus, the resulting $d\Gamma/dq^2$ distribution can be used as a test of the form factor calculations.  

\begin{figure}
\hbox to\hsize{\hss
\includegraphics[width=\hsize]{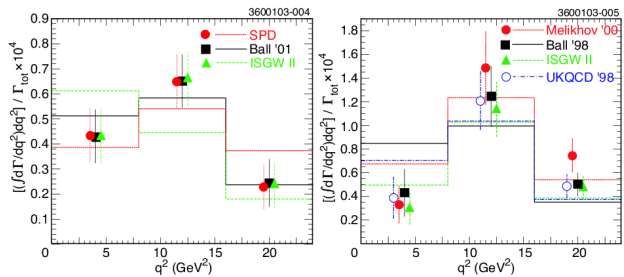}
\hss}
\caption{The $d\Gamma/dq^2$ distributions obtained in the CLEO '03 analysis for $B^0\to\pilnu{-}$ (left) and $B^0\to\rholnu{-}$ (right).  Shown are the variations in the extracted rates (points) for form factor calculations that have significant $q^2$ variations, and the best fit of those shapes to the extracted rates (histograms).}
\label{fig:dgamma}
\end{figure}

The CLEO '03 analysis adopts both approaches, and the resulting $q^2$ distributions are shown in Figure~\ref{fig:dgamma}, along with a variety of calculations.  The $\pilnu{\ }$ mode shows essentially no variation as the form factors are changed.  The $\rholnu{\ }$ mode still exhibits dependence on the form factor calculation.  The branching fractions, obtained by summing the measured rates, show significantly smaller form factor dependence than other measurements (Figure~\ref{fig:branching_fractions}) for both modes.  The residual uncertainty in the $\rholnu{\ }$ mode likely results from a cut on the angle between the lepton in the $W$ rest frame and the $W$ direction and the different effect this has on the three form factors for that decay. 

Cross-feed among $\btou$ modes also exhibits a strong $q^2$ dependence, with the high--$q^2$ regions seeing the most background contribution.  This phenomenon is clearly visible in the cross-feed background contribution to $\pilnu{\ }$ in Figure~\ref{fig:cleo_pi}.  Basically, at large $q^2$, which corresponds to soft mesons, there is a larger probability for random hadronic combinatorics to combine with the lepton pair from a non-signal $\btou$ mode and satisfy the kinematic requirements for the signal mode.  Most analyses have not studied variations of the form factors from specific background modes.  The CLEO '03 approach of extracting rates in independent $q^2$ intervals has reduced the sensitivity to the form factors relative to the CLEO '96 analysis.
 
\homework{What set of form factor calculations should all experiments use to evaluate the uncertainty of the signal rates on the signal form factors?}
\homework{Similarly, what sets should be used to evaluate cross-feed background from $\pilnu{\ }$, $\rholnu{\ }$, or $\omelnu$ into the signal $\btou$ mode?}
\homework{Given the observed variations in rates, how should the uncertainties be assigned?}
\homework{Given the uncertainties, how should they be combined with other uncertainties in an averaging procedure?}

Finally,  we must consider evaluation of the systematics associated with simulation of generic $\btou$ backgrounds.  In principle, these uncertainties will be strongly correlated between analyses, but the broad variation in approaches to modeling the decays and evaluating the systematics make analysis of the correlations currently impossible.  Unfortunately, these systematics make sizable contributions to the measurement of the vector modes in particular.
The experiments should standardize on various aspects of simulation and systematic testing.  

\homework{What exclusive modes and inclusive parameters should be used in simulation, and how does one ``marry'' the exclusive and inclusive decays?}
\homework{What variation on exclusive modes and form factors and on inclusive parameters should be made?}
\homework{How should the ``inclusive'' decays be hadronized, and how should the hadronization be varied?}
\homework{How should potential contributions from nonresonant $\pi\pi\ell\nu$ (or other nonresonant final states) be evaluated, and potential correlations between experiments probed?} 

\subsection{Extraction of $|V_{ub}|$}

\begin{figure}
\hbox to\hsize{\hss
\includegraphics[width=\hsize]{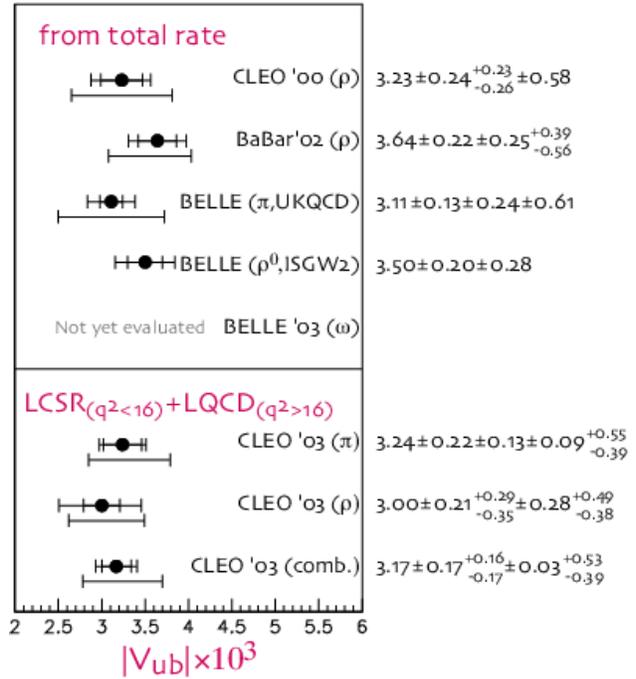}
\hss}
\caption{$\Vub$ results obtained from the different analyses reviewed here. Top: values obtained using the entire $q^2$ range.  Bottom: values obtained using only LCSR and LQCD calculations, restricted to the intervals $q^2<16$ GeV$^2$ and $q^2>16$ GeV$^2$ to minimize modeling uncertainties that are difficult to quantify.}
\label{fig:vub_results}
\end{figure}

Two methods for extraction of $|V_{ub}|$ have been used in the analyses.  The first, used in most analyses, simply takes the measured partial rate for the signal mode obtained using a given form factor calculation and extracts $\Vub$ using that calculation's predicted $\Gamma/\Vub^2$.  The resulting values for $\Vub$ are summarized in the top portion of Figure~\ref{fig:vub_results}.  Note that the quoted theoretical uncertainties on $\Vub$ obtained by the experiments are {\em NOT} directly comparable.  They are evaluated with rather different degrees of conservativeness and analyses with similar sensitivities will appear with differing uncertainties.

Independent extraction of rates in the three $q^2$ intervals in the CLEO '03 analysis permits extraction of $\Vub$ with a method that minimizes unquantifiable modeling assumptions.
The CLEO '03 analysis restricts consideration to those calculations based firmly in QCD: LQCD and LCSR.  However, extrapolation of those calculations outside of the $q^2$ range in which the techniques are valid to the full $q^2$ range involves modeling assumptions that cannot be quantified.  CLEO therefore uses the measured rates only in the ranges of validity for each technique, thereby minimizing the modeling assumptions.  Those results are also summarized in Figure~\ref{fig:vub_results}.

To provide a test of the various different form factor predictions over the full $q^2$ range,
the CLEO '03 analysis also fits the predicted $d\Gamma/dq^2$ from a particular calculation to the measured distribution obtained with that calculation.  To obtain the optimal normalization, $\Vub$ is allowed to float.  The quality of each fit then provides a test of each model.  For $\pilnu{\ }$, the fit quality for ISGW~II is considerably poorer than for the other calculations tested (see Figure~\ref{fig:pi_models}).  
Note that while similar results with similar numerical uncertainties for $\Vub$ are obtained with this
technique.  Because the results rely on more modeling assumptions and are therefore less robust
than the other procedure, CLEO does not take these values as the primary results.

\section{Summary}

We are entering an exciting phase for measurement of exclusive $\btou$ branching fractions and extraction of $\Vub$.  Many new measurements with new techniques have been presented recently, with more anticipated.  The statistics are now sufficient to permit extraction of $d\Gamma/dq^2$ with reasonable precision in the decays, which in turn allows reduction of the uncertainty from theoretical form factor shapes on the rates.  Measurement of $d\Gamma/dq^2$ also allows more robust extraction of $\Vub$.  

The theoretical prediction on the overall form factor normalization continues to dominate extraction of $\Vub$ from the measured rates.  I wait with great anticipation for the first unquenched LQCD calculations over a much broader range of $q^2$.

\end{document}